\documentclass[conference]{IEEEtran}
\usepackage[utf8]{inputenc}
\usepackage[T1]{fontenc}
\usepackage{mathptmx}  % Type1 Times: T1-safe glyphs AND correct text extraction
\usepackage{amsmath,amssymb}
\usepackage{booktabs}
\usepackage{graphicx}
\usepackage{xcolor}
\usepackage{url}
\usepackage[hidelinks]{hyperref}

\title{What Actually Serializes GPU LZ77 Decode\\
\large Three decoders, three mechanisms, and an encode-time lever that
removes the last one}

\author{\IEEEauthorblockN{Yakiv Shavidze}
\IEEEauthorblockA{Independent Researcher\\
ORCID: 0009-0008-3622-3448\\
\url{https://github.com/yasha1971-coder/aceapex}}}

\begin{document}
\maketitle

\begin{abstract}
The sequential part of GPU LZ77 decode is not where the field assumes it is.
Across three decoder architectures on an H100 we measure that parse, not
copy, holds 64--72\% of device-resident decode time; that bounding
back-reference chain depth---provable, and costing 0.006\% in ratio---moves
latency by at most 2.8\% and, for the file's own latency spike, provably by
nothing at all, since a byte-level comparison of all 15{,}499 blocks shows
the cap alters none of the 181 blocks involved; that self-overlapping
matches are periodic fills rather than dependency chains, which makes them
fully parallel and speeds the match layer by 2.75--8.42$\times$ bit-perfect;
and that the last genuinely sequential element, a four-entry distance
history, can be removed by the encoder for 0.540\% of ratio, growing the
dependency-free parse run from 4 commands to 706. We also report the floor
the format runs into: with a median match of 7 bytes against a 128-byte
cache line, bus efficiency is 4.4\% and a coalesced write of the same data
is 39$\times$ faster. A separate section records ten hypotheses these
measurements refuted, including one methodological error of our own. Every
reproducible claim carries a machine-checkable record: a fresh clone of the
tagged release passes 17 of 17 checks reachable without a GPU, none failing.
\end{abstract}


\section{Introduction}
Consider the situation directly. You have a 50\,GB genome, 80\,GB of VRAM,
and a request for a 16\,KB region. Decompressing everything leaves nowhere
to put it. Holding the data compressed and decoding on demand is possible,
but then the response time is set by something you do not control. The
engineer's question is simple: \emph{what exactly is sequential in a GPU
decode, and can it be removed?}

The field's answer so far has been ``back-reference dependencies.''
Gompresso~\cite{gompresso} eliminates them at compression time by
restricting matches to a per-warp high-water mark, at up to 19\%
degradation in compression ratio. We measured where the sequential part
actually lives, and removed a different one---in the parse---for 0.540\%.

This paper is a measurement study, and its structure follows what the
measurements found rather than what we expected to find. Three decoder
architectures are used deliberately, because each exposes a different
limiting mechanism, and a claim that survives only one of them is not a
claim about the format.

\paragraph{Three questions this answers}
\begin{itemize}
\item \emph{What does random access into compressed data cost?} The cost is
blockwise encoding: 6.7\% on genome, 24.2\% on text---and the match layer
returns more than half of it.
\item \emph{Why does my GPU decoder achieve a few percent of bandwidth?}
The median match is 7 bytes against a 128-byte cache line; bus efficiency is
4.4\% and the coalesced ceiling is 39$\times$ away.
\item \emph{What should be optimized?} Parse, at 64--72\% of decode time.
Not copy, and not chain depth.
\end{itemize}

\section{Setup}
ACEAPEX stores every back-reference as an absolute position in the
decompressed output rather than a relative distance in a sliding window. The
encoder performs a global match search and partitions the output into
fixed-size blocks; because offsets are absolute, a block decodes as soon as
the blocks holding its sources are present. Four streams are stored per
block: literals, lengths, absolute offsets, and commands.

Three decoders are used throughout, and Table~\ref{tab:decoders} states what
governs time in each. Confusing them is the main way to misread the numbers
that follow.

\begin{table}[t]
\caption{Three decoder architectures and what sets their time.}
\label{tab:decoders}
\centering
\small
\begin{tabular}{@{}lp{2.6cm}p{2.7cm}@{}}
\toprule
decoder & scheme & time governed by \\
\midrule
dense full-pipe & ANS + one match kernel & parse \\
wavefront & one launch per level, CUDA graph & work $+\;N_\text{waves}\times4.5\,\mu$s \\
v7-RA seek & persistent queue, leader per block & token count \\
\bottomrule
\end{tabular}
\end{table}

All measurements are on one H100 80\,GB SXM with an EPYC 4344P host, CUDA
12.x. The canonical corpus is human chromosome~1 (UCSC hg38,
253{,}935{,}557\,B, 16\,KB blocks, 15{,}499 blocks). Correctness is
bit-perfect by FNV or XXH3 against the original bytes; the few numbers
without that check are marked where they appear.

\paragraph{Bench noise} Repeated dense full-pipe runs on chr1 give
69.1 / 73.1 / 69.0\,GB/s. We treat \textbf{6\%} as the resolution of this
bench and do not claim effects below it.

\section{The Parse Wall}
\begin{figure}[t]
\centering
\includegraphics[width=\columnwidth]{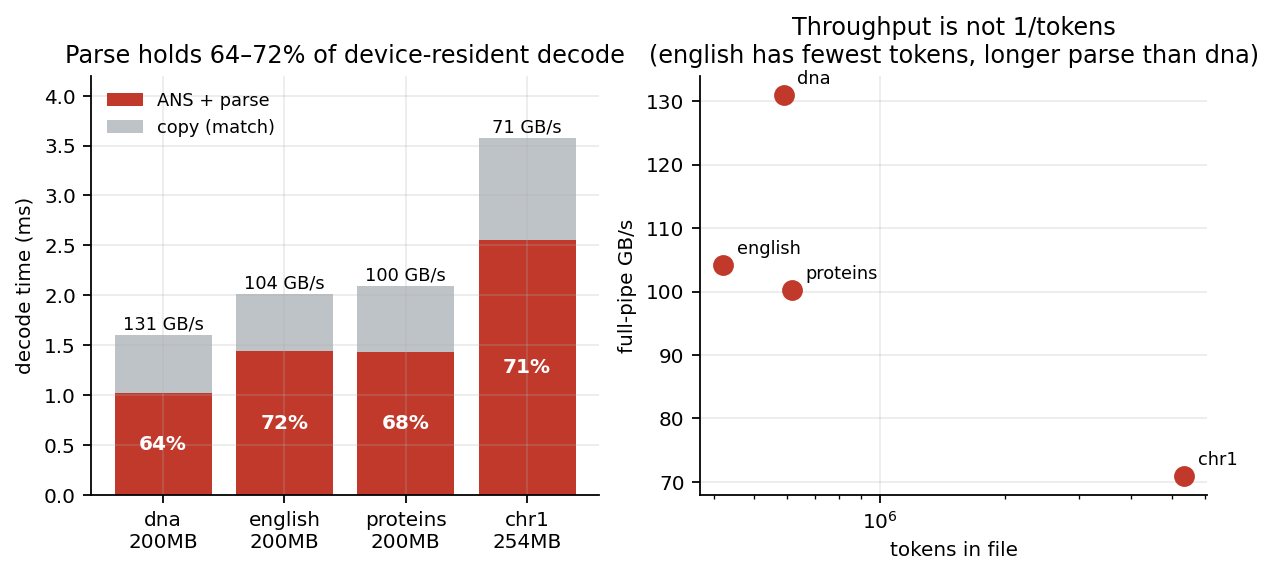}
\caption{Parse holds 64--72\% of device-resident decode on four corpora.
Throughput is not proportional to $1/\text{tokens}$: english has the fewest
tokens and a longer parse than dna.}
\label{fig:parse}
\end{figure}

Table~\ref{tab:parse} and Figure~\ref{fig:parse} give the split of
device-resident decode across four corpora. Parse holds 63.7--71.5\% of the
time in every case. This is the paper's main quantitative contribution and
it closes an open item of the series: earlier work in this line reported an
anchor of roughly 50\% measured on a single corpus.

\begin{table}[t]
\caption{Full device-resident decode. Parse dominates on every corpus.
All bit-perfect.}
\label{tab:parse}
\centering
\small
\begin{tabular}{lrrrrr}
\toprule
corpus & tokens & ANS+parse & copy & parse & GB/s \\
\midrule
dna 200\,MB      & 589{,}469   & 1.020\,ms & 0.581 & 63.7\% & 131.0 \\
english 200\,MB  & 421{,}940   & 1.440     & 0.573 & 71.5\% & 104.2 \\
proteins 200\,MB & 617{,}836   & 1.432     & 0.659 & 68.5\% & 100.3 \\
chr1 254\,MB     & 5{,}365{,}972 & 2.555   & 1.021 & 71.5\% & 71.0 \\
\bottomrule
\end{tabular}
\end{table}

The same table refutes a natural model. If throughput were governed by token
count alone, english---with the fewest tokens---would have the shortest
parse. It has a longer one than dna.

\section{Depth Does Not Govern Latency}
The depth cap works, and it does not do what we expected. We report both
halves.

\subsection{The bound is real and nearly free}
A two-pass, forced-literal encoder holds every block's chain depth at or
below a chosen $L$. The first pass encodes, decodes its own output and
derives each token's depth; matches deeper than $L$ contribute their span to
a forced-literal set; the second pass rejects matches covering forced
regions. Re-parsing creates new deep chains, so the process iterates until
it converges. The container is unchanged---only match selection differs, so
any $L$ is read by the same decoder.

Table~\ref{tab:cost} gives the cost on the full corpus. It is within
$\pm0.006\%$ and negative at $L{=}32$. Before capping, 135 / 37 / 16 blocks
exceed $L{=}16$ / 24 / 32; after capping, none do.

\begin{table}[t]
\caption{Cost of the depth guarantee. Full chr1, baseline ratio 3.18065,
all bit-perfect.}
\label{tab:cost}
\centering
\small
\begin{tabular}{lrrr}
\toprule
cap $L$ & max depth & ratio & cost \\
\midrule
64 (natural) & 50 & 3.18065 & $0.0000\%$ \\
32 & 32 & 3.18078 & $-0.0041\%$ \\
24 & 24 & 3.18066 & $-0.0003\%$ \\
16 & 16 & 3.18048 & $+0.0053\%$ \\
\bottomrule
\end{tabular}
\end{table}

\begin{figure}[t]
\centering
\includegraphics[width=\columnwidth]{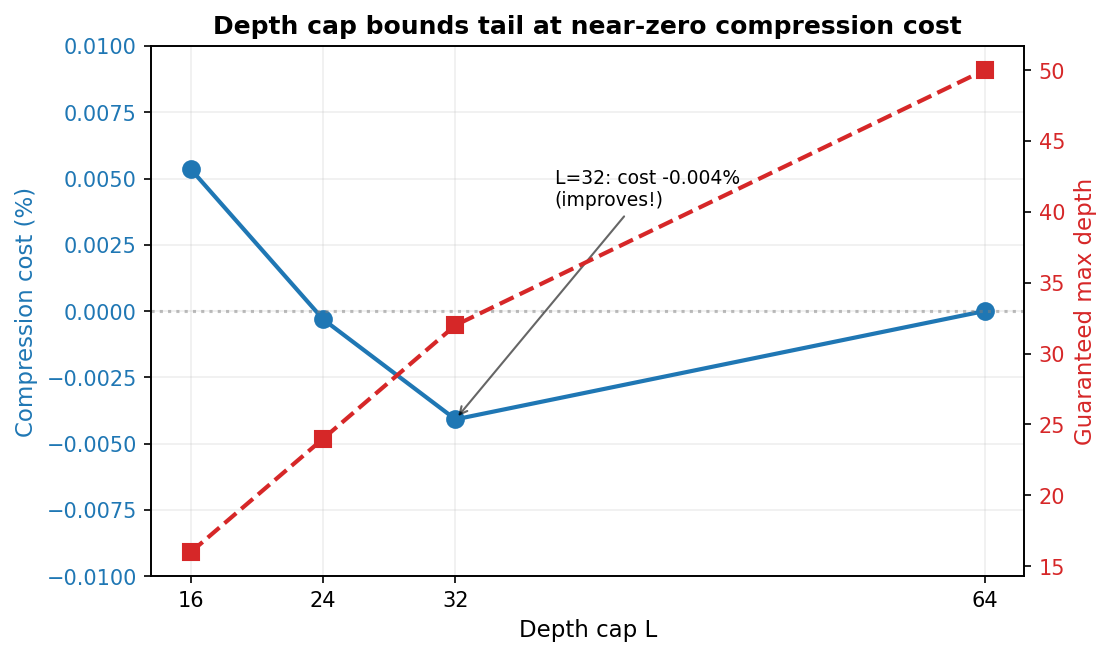}
\caption{Tightening the cap lowers guaranteed maximum depth at a compression
cost within $\pm0.006\%$; at $L{=}32$ the cost is negative.}
\label{fig:cost}
\end{figure}

\subsection{What the cap does to time}
On the wavefront decoder, where each level is a kernel launch, depth is
visible and small: 51 waves at $L{=}\infty$ down to 17 at $L{=}16$, for
$-2.63\%$ (5.846 to 5.692\,ms, noise 0.003\,ms). The marginal cost of a wave
is $154\,\mu\text{s}/34 = 4.5\,\mu$s; repeat baseline readings
(5.846--5.852\,ms) put it at 4.5--4.7\,$\mu$s.

On the dense full-pipe the effect disappears below bench noise: the ranges
at $L{=}50$, 32 and 16 overlap. One signal is stable---copy time at $L{=}16$
is 0.965--0.969\,ms against 1.022--1.040 ($-6\%$, non-overlapping), because
the cap literalizes deep matches and literals are written coalesced---but
parse is 70\% of the total, so the effect is submerged.

The decisive experiment is on v7-RA seek, at block 7700, with three
configurations of \emph{equal token mass} (spread 295 tokens out of 5.3
million, 0.0055\%). Table~\ref{tab:abc} shows that only the configuration
which actually reduces cluster depth moves the time; forcing an equal number
of leaf tokens, or an equal number outside the cluster, changes nothing.
FNV is identical across all four.

\begin{table}[t]
\caption{Equal-token-mass controls at block 7700. Only reducing depth (A)
moves latency; leaf (B) and placebo (C) do not.}
\label{tab:abc}
\centering
\small
\begin{tabular}{lrrr}
\toprule
config & $\Delta$tokens & cluster depth & $\mu$s \\
\midrule
baseline & --- & 4.358 & 480.4 \\
A (deep forced) & $-20{,}837$ & 2.615 & 467.1 \\
B (leaf forced) & $-20{,}668$ & 4.542 & 480.5 \\
C (placebo, outside) & $-20{,}963$ & 4.358 & 480.5 \\
\bottomrule
\end{tabular}
\end{table}

\begin{figure}[t]
\centering
\includegraphics[width=\columnwidth]{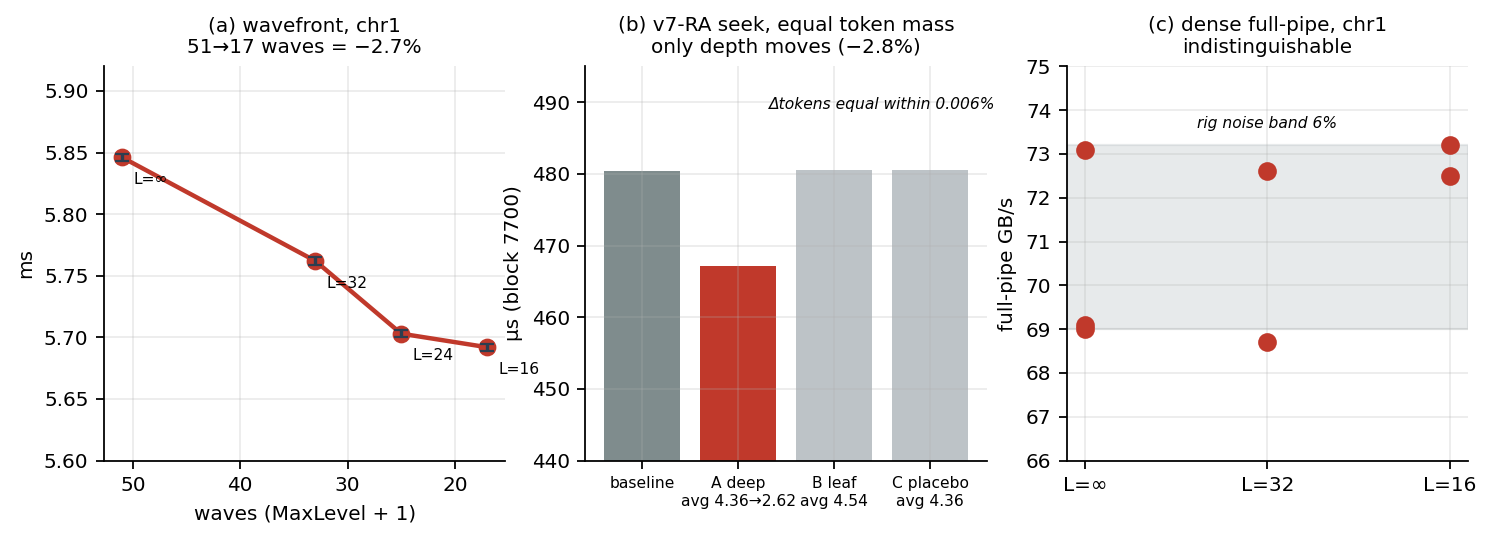}
\caption{Depth across three decoders: visible and small on wavefront
($-2.6\%$), isolated by placebo control on v7-RA seek ($-2.8\%$ for the deep
configuration only), and indistinguishable from bench noise on the dense
full pipe.}
\label{fig:three}
\end{figure}

\subsection{The byte-level refutation}
\begin{figure}[t]
\centering
\includegraphics[width=\columnwidth]{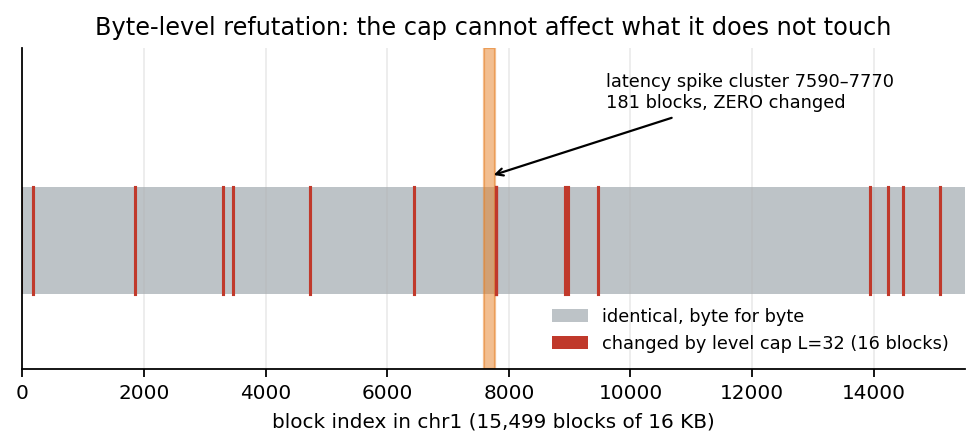}
\caption{Byte-level comparison of all 15{,}499 blocks. The cap changes 16
blocks, and none of them lie in the latency spike cluster.}
\label{fig:hash}
\end{figure}

The strongest statement needs no GPU at all. Comparing all four streams of
every one of the 15{,}499 blocks, baseline against $L{=}32$, exactly
\textbf{16 blocks} differ, and that set coincides exactly with the
forced list. The file's latency spike cluster, blocks 7590--7770, contains
\textbf{none of them}: 0 of 181 blocks changed, byte for byte
(Figure~\ref{fig:hash}).

A cap that alters no byte in a region cannot alter that region's decode
time. This is a negative result carrying its own positive control.

\subsection{A reconciling model}
The three decoders agree under
$\text{latency} \approx a\cdot\text{waves} + b\cdot\text{bytes}$, with
$a = 4.5$--$4.7\,\mu$s---the price of an empty wave---and
$b$ dominant, ranging 30--181\,$\mu$s/MB across corpora. For the spike
cluster the wave term is identically zero, which is why depth has no purchase
there.

The spike itself is explained by parse, like everything else. A cluster block
issues 772 references to 99 distinct cache lines (a 12.7\,KB working set,
L1-resident) against 473 references to 119 lines in a control block; it is
not memory-pathological but \emph{instruction-dense}, at 767 tokens against a
median 357. Its latency ratio, 493 against 386\,$\mu$s, is smaller than the
token ratio because a single-block seek on v7-RA is launch-bound. Solving
$386 = F + c\cdot357$ and $493 = F + c\cdot767$ gives a fixed cost
$F = 293\,\mu$s and $c = 0.26\,\mu$s per token; the fixed part agrees with
the 292--387\,$\mu$s single-tile figures of Section~\ref{sec:seek50},
measured on a different corpus. Only the remainder scales with tokens.

\begin{figure}[t]
\centering
\includegraphics[width=\columnwidth]{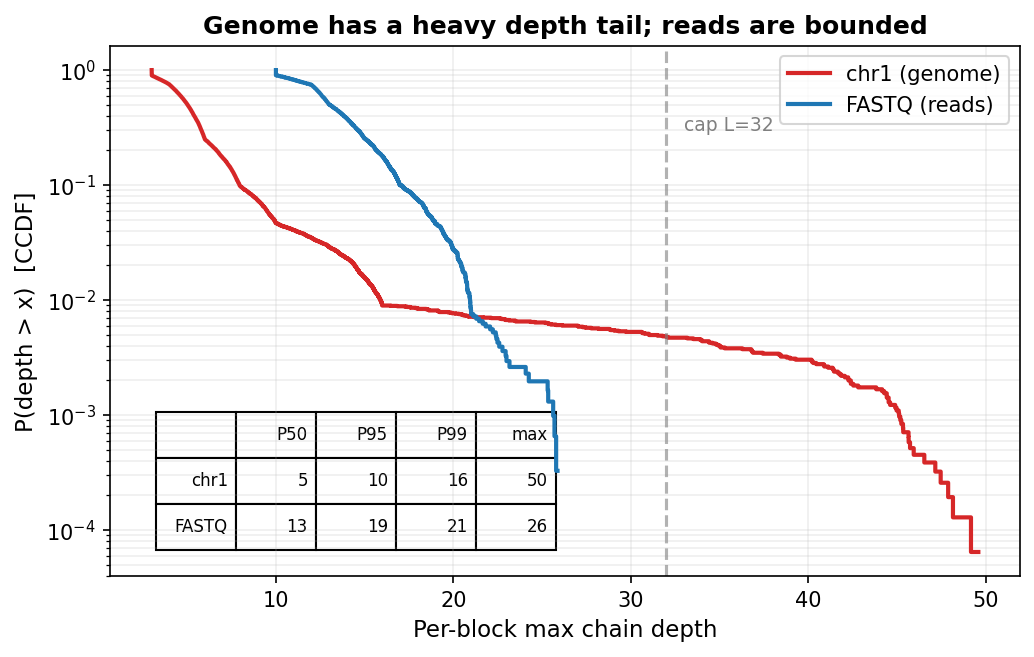}
\caption{The genome carries a heavy depth tail; reads are bounded. The curves
cross near depth 21.}
\label{fig:ccdf}
\end{figure}

\begin{figure}[t]
\centering
\includegraphics[width=0.84\columnwidth]{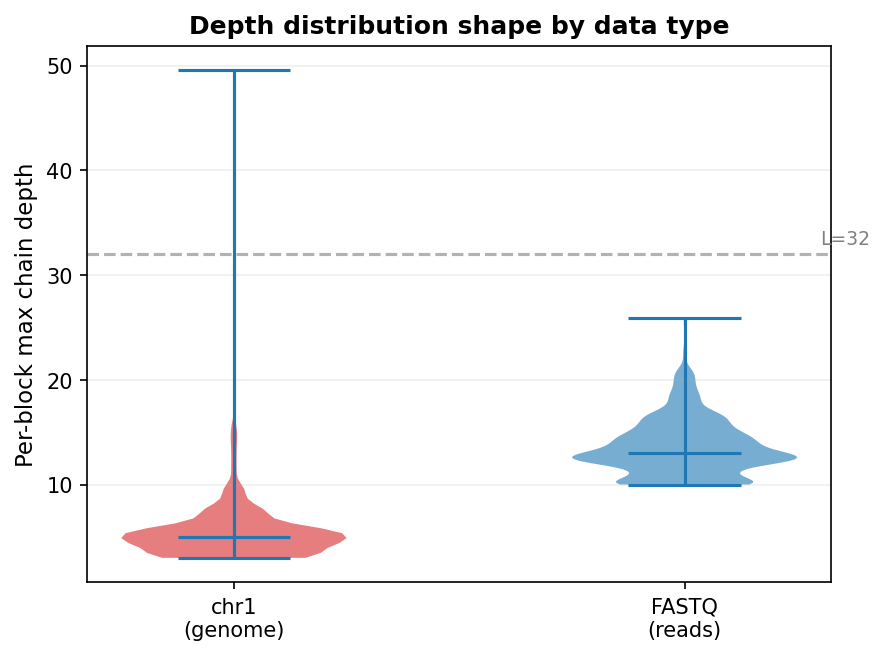}
\caption{Two risk profiles: a narrow body with a long whisker to 50 (genome)
against a thick moderate body ending at 26 (reads).}
\label{fig:violin}
\end{figure}

Where the depth tail lives at all is a property of the data. A genomic FASTA
corpus has a light body with rare very deep blocks (median max-depth 5,
maximum 50); a read corpus is uniformly moderate (median 13, maximum 26) and
has no extreme to cut (Figures~\ref{fig:ccdf} and~\ref{fig:violin}).

\section{What Does: Periods, Not Chains}
A self-overlapping match, where the source range intersects the destination,
is treated by every implementation we know as inherently serial. It is not.
Writing the fill as $\texttt{out}[dst+k] = \texttt{out}[src + k \bmod dist]$
places the source entirely \emph{outside} the written range, so the threads
are independent. The period $dist$ is known to the encoder from the start---it
is in the format.

On chr1 there are 450{,}984 self-overlapping tokens with $dist$ at
P50 = 3, P90 = 5, P99 = 9, maximum 153 (microsatellites). Take the critical
path to be the sum over levels of the longest match at that level, since a
wave cannot finish before its longest match does: a single level-2 match of
length 16{,}333 at period $\approx3$ accounts for \textbf{57\%} of it
(16{,}333 of 28{,}587).

The patch has two parts and they work only together: the modular semantics,
which permits parallelism, and one warp per token, which uses it. Semantics
alone on a single thread gives $-2.4\%$; with the warp mapping, chr1 falls
from 5.846 to 1.291\,ms ($-77.9\%$). Table~\ref{tab:period} gives the four
corpora; all bit-perfect, and chr1 repeats at 1.291 three times.

\begin{table}[t]
\caption{Match layer with period-aware fill and one warp per token.
Match layer only---entropy is outside this timer.}
\label{tab:period}
\centering
\small
\begin{tabular}{lrrr}
\toprule
corpus & before (ms) & after & speedup \\
\midrule
proteins & 5.571 & 0.662 & $8.42\times$ \\
chr1 & 5.846 & 1.291 & $4.53\times$ \\
silesia & 16.018 & 5.346 & $3.00\times$ \\
enwik9 & 7.804 & 2.838 & $2.75\times$ \\
\bottomrule
\end{tabular}
\end{table}

\begin{figure}[t]
\centering
\includegraphics[width=\columnwidth]{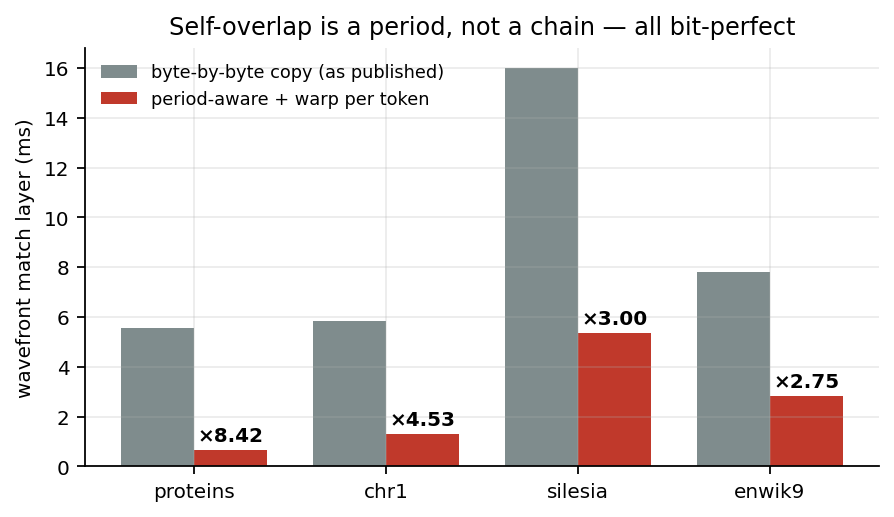}
\caption{Self-overlap is a period, not a chain: $2.75$--$8.42\times$ on the
match layer, bit-perfect.}
\label{fig:period}
\end{figure}

\section{The Last Sequential Element}
\label{sec:lastseq}
Decomposing all 10{,}566{,}105 commands of chr1 shows what is genuinely
sequential: the four-entry distance history behind repeat matches accounts
for 5.45\% of commands, varint lengths for 0.61\%, and the remaining
\textbf{93.94\%} is parallelisable by prefix sum.

That is demonstrated rather than argued. A pure prefix sum carrying no state
reproduced every position of the sequential parse: block 1000 (761 commands)
and block 15000 (746) show \emph{zero} discrepancies in both output and
literal positions. Block 5000 shows 30 discrepancies with exactly one varint
command; block 0 shows 335 at 2.4\% varint. Discrepancies are strictly
proportional to varint expansion.

Removing the history is an encoder-side decision. With repeat codes
suppressed---the distance goes out as a varint, the match itself is
preserved, the format is unchanged and the decoder is the same---the chained
share falls from 5.46\% to 0.02\% and the dependency-free run grows from 4
commands to 706 at the median, a factor of 176. The cost is
\textbf{0.540\%} of ratio (3.18065 to 3.16347)---equivalently 0.543\% growth
in compressed size, which is the figure the reproduction record reports; the
control, with the flag off, returns exactly 3.18065. Both are bit-perfect.

The comparison with prior work needs care, because the object differs.
Gompresso's dependency elimination removes a \emph{match-layer} dependency
(nested references inside a warp); the element removed here is a
\emph{parse-layer} one---absent from DEFLATE-style formats and present in
zstd. The costs are not directly comparable, but both are encode-time
decisions with a price in ratio: theirs up to 19\%, ours 0.540\%. The
match-layer dependency itself is what absolute offsets and block
independence already remove in this format~\cite{aceapex1}.

\begin{figure}[t]
\centering
\includegraphics[width=\columnwidth]{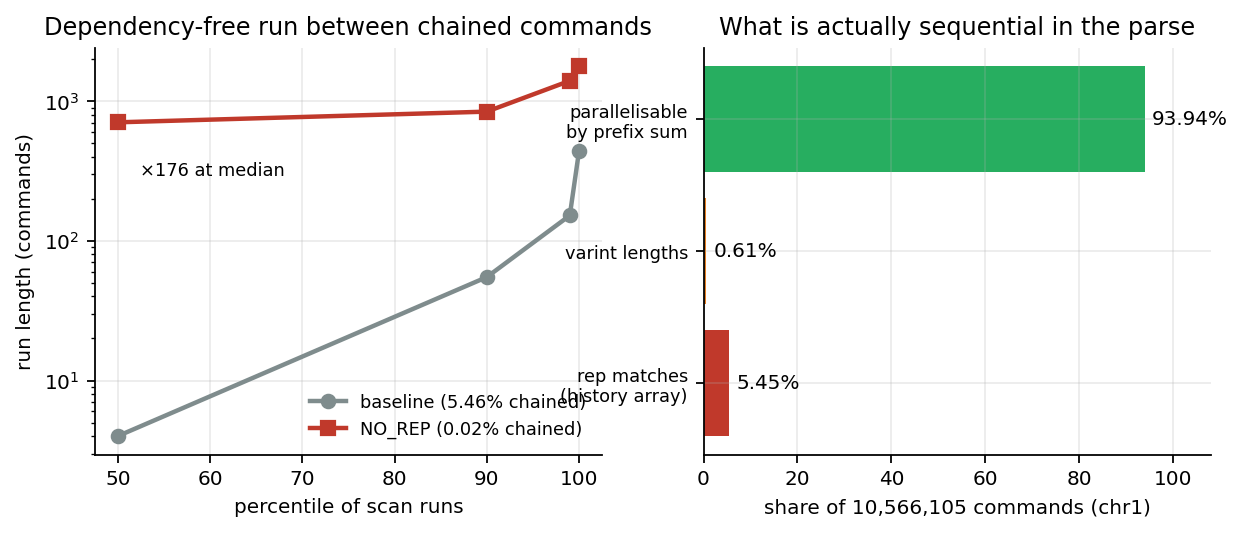}
\caption{The dependency-free run between chained commands grows
$176\times$ at the median; 93.94\% of commands are already parallelisable by
prefix sum.}
\label{fig:runs}
\end{figure}

\section{The Memory Floor}
\label{sec:floormem}
What remains after the sequential part is removed is a bus problem.
Copying 61{,}089{,}878 logical bytes on chr1 moves 1{,}392{,}400{,}000 bytes
of cache-line traffic at 128\,B granularity: \textbf{4.4\%} bus efficiency.
An independent check writes the same 61\,MB coalesced in 0.027\,ms
(2250.3\,GB/s) against our 1.056\,ms (58\,GB/s)---a gap of $39\times$.
Writes are 82\% of the match-layer time, source reads 18\%.

\begin{figure}[t]
\centering
\includegraphics[width=\columnwidth]{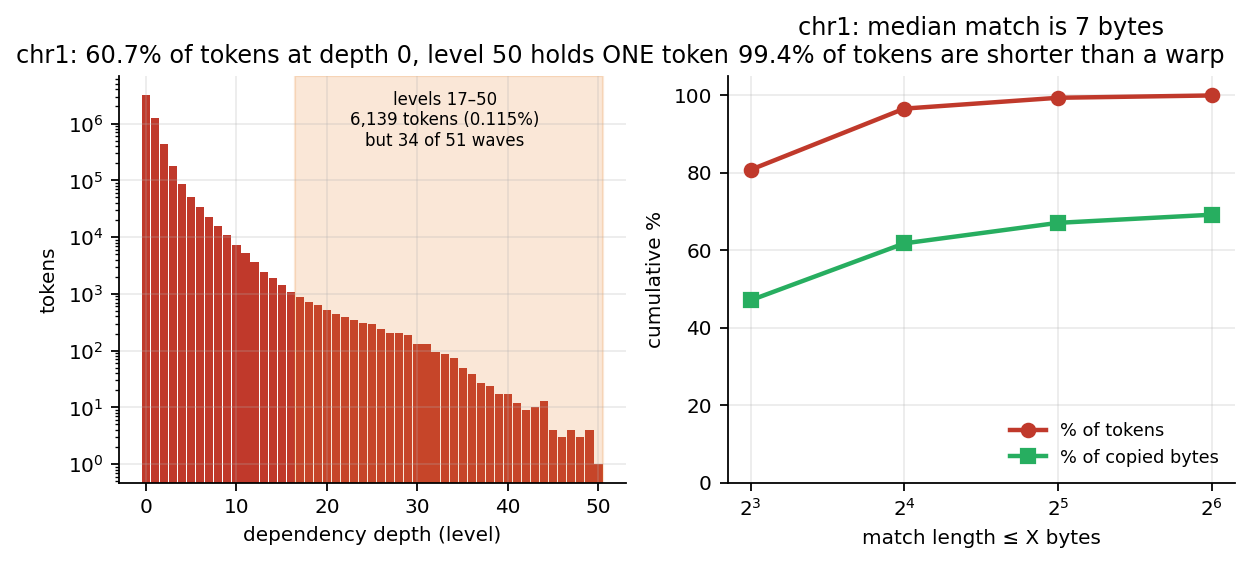}
\caption{Depth histogram and match-length CDF for chr1: 60.7\% of tokens at
depth 0, and a median match of 7 bytes.}
\label{fig:dist}
\end{figure}

The cause is granularity: the median match is 7 bytes and 99.4\% of tokens
are shorter than a warp (Figure~\ref{fig:dist}). This independently
rediscovers the work-granularity mechanism reported in Paper~4 of this
series~\cite{aceapex4}, from a different direction. Consistent with that
paper, sorting tokens by destination within a level changes nothing
(1.256 against 1.291\,ms, noise): address scatter is not the bottleneck.

\begin{figure}[t]
\centering
\includegraphics[width=\columnwidth]{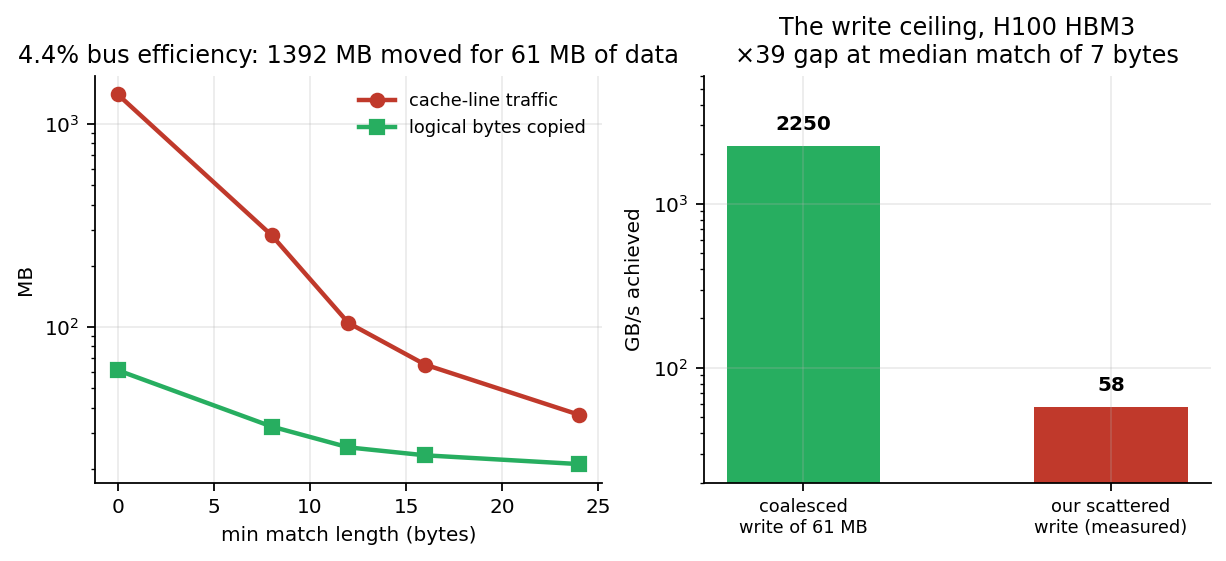}
\caption{4.4\% bus efficiency: 1392\,MB of traffic for 61\,MB of data, and a
$39\times$ gap to a coalesced write.}
\label{fig:bus}
\end{figure}

\section{Random Access at Scale}
\label{sec:seek50}
On a 50{,}000{,}000{,}000-byte archive of 3{,}051{,}758 blocks
(22.8\,GB of streams, 50\,GB of output, 72.8 of 79.2\,GB VRAM), decoding one
16\,KB tile takes 335.6, 386.6, 360.5 and 292.6\,$\mu$s at the start, at
16\,GB, at 33\,GB and at the end---a spread of $\pm14\%$ around 344\,$\mu$s
with four distinct FNV values and no trend with position. Widening the
request scales far better than linearly: one block in 0.339\,ms, one thousand
blocks (16\,MB) in 0.423\,ms.

\begin{figure}[t]
\centering
\includegraphics[width=\columnwidth]{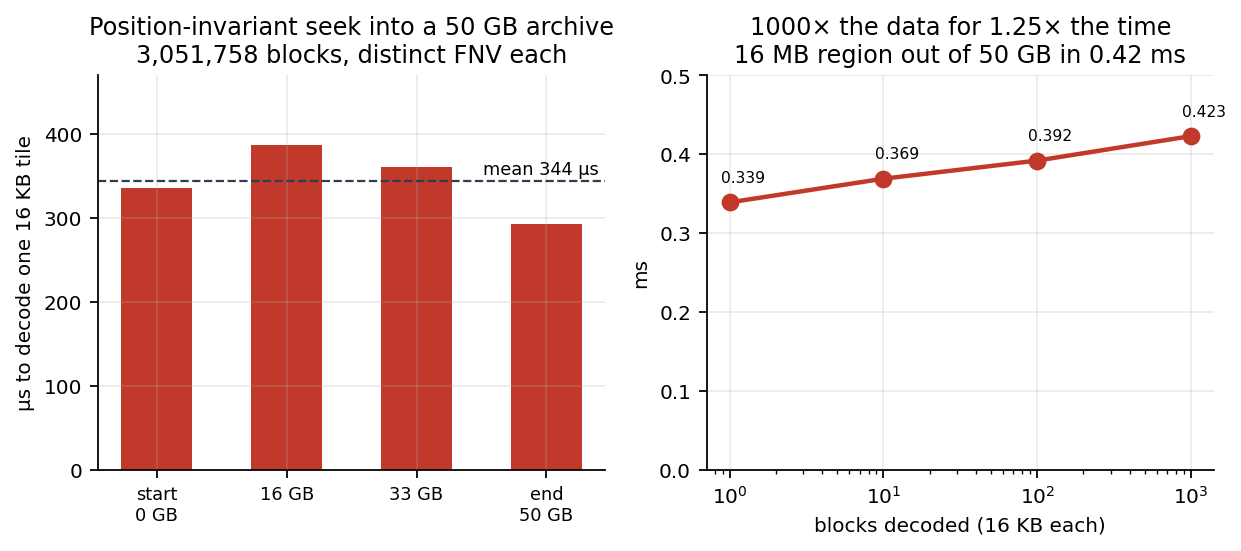}
\caption{Position-invariant seek into a 50\,GB archive, and a
$1000\times$ larger region for $1.25\times$ the time.}
\label{fig:seek}
\end{figure}

We state the boundary: the 50\,GB original is not on disk, so there is
nothing to compare the FNV values against. This is position-invariance at
50\,GB; bit-perfect correctness is shown on the smaller corpora.

\section{The Class}
Comparing a random-access format against a whole-stream archive measures the
constraint, not the mechanism. Table~\ref{tab:class} therefore places every
codec under the same 16{,}384-byte independent-block constraint, so all are
equally seekable.

\begin{table}[t]
\caption{Compression ratio with independent 16\,KB blocks for all codecs.}
\label{tab:class}
\centering
\small
\begin{tabular}{lrrrrrr}
\toprule
corpus & lz4 & zstd-3 & brotli-9 & zstd-19 & ACEAPEX & rank \\
\midrule
chr1 & 1.786 & 3.034 & 3.287 & \textbf{3.475} & 3.181 & 3 \\
enwik8 & 1.595 & 2.334 & 2.564 & 2.488 & \textbf{2.638} & 1 \\
enwik9 & 1.750 & 2.567 & 2.824 & 2.736 & \textbf{2.981} & 1 \\
silesia & 1.907 & 2.675 & 2.882 & 2.936 & \textbf{3.005} & 1 \\
FASTQ 1\,GB & 2.324 & 3.629 & 3.910 & \textbf{4.029} & 3.965 & 2 \\
\bottomrule
\end{tabular}
\end{table}

\begin{figure}[t]
\centering
\includegraphics[width=\columnwidth]{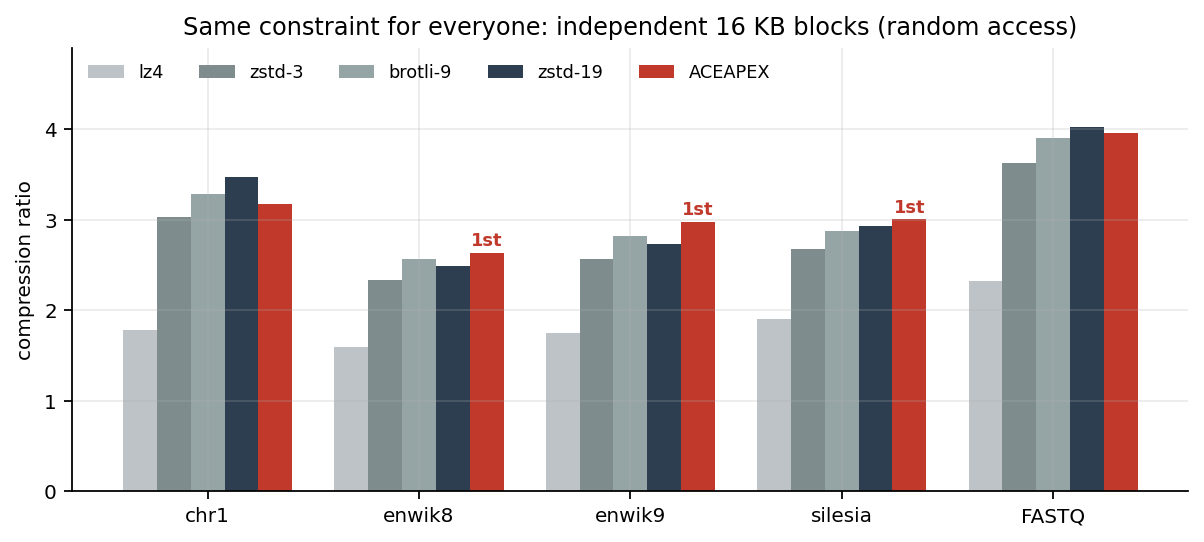}
\caption{Under an equal 16\,KB independent-block constraint, ACEAPEX ranks
first on three of five corpora against zstd-19.}
\label{fig:class}
\end{figure}

At equal seekability ACEAPEX ranks first on three of five corpora against
zstd-19, and ahead of the speed-comparable zstd-3 everywhere, by 4.9\%
(chr1) to 16.1\% (enwik9). The price of addressability is the blockwise
constraint itself: whole-stream zstd-3 against blockwise zstd-3 differs by
6.68\% on chr1 and 24.19\% on enwik9---and the match layer returns nearly
three quarters of it (3.034 to 3.181 on chr1, recovering 4.85 of the 6.68).

\section{What We Refuted}
Every entry below is a measurement, not an opinion. We include our own
methodological error.

\begin{figure}[t]
\centering
\includegraphics[width=\columnwidth]{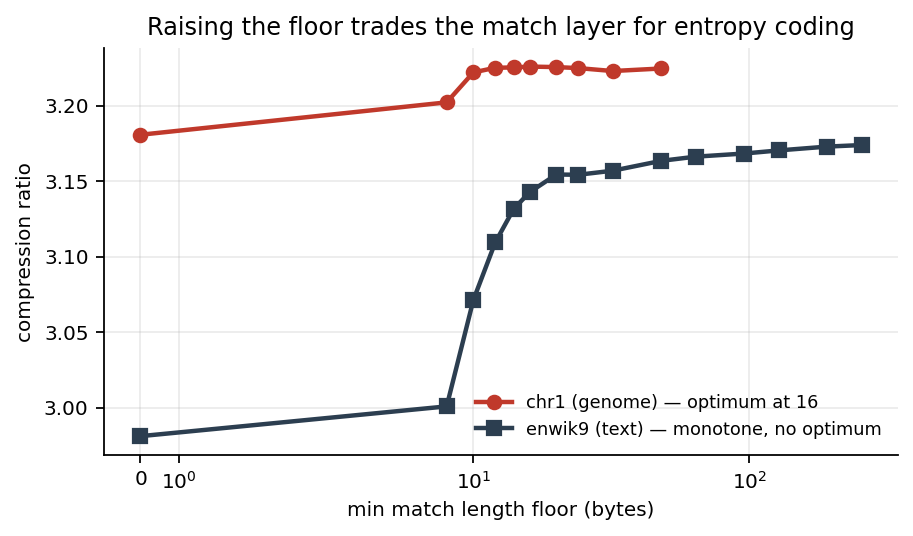}
\caption{Raising the minimum match length trades the match layer for entropy
coding, with an optimum on genome and none on text.}
\label{fig:minmatch}
\end{figure}

\begin{enumerate}\itemsep2pt
\item \emph{The cap cuts the spike cluster.} Zero overlap at $L{=}32$; 0 of
181 blocks changed byte for byte.
\item \emph{Latency scales with maximum level.} Counterexample: english
(58 waves, 200\,MB) is $6.3\times$ faster than proteins (45 waves, 200\,MB).
\item \emph{A 4.7\,$\mu$s-per-wave slope is universal.} It ranges 7.8--123.3
across six corpora ($16\times$). The 4.7 figure is the price of an
\emph{empty} wave; the average on chr1 is 114.6.
\item \emph{Throughput is proportional to $1/\text{tokens}$.} English has
the fewest tokens and a longer parse than dna.
\item \emph{A scan parser helps on the current format.} The run between
chained commands is 4 commands at the median; a 32-thread warp on 4 commands
loses to sequential.
\item \emph{Sorting tokens by destination will coalesce writes.} Mean
destination stride at level 2 is 578\,B before and 578\,B after---already
sorted; 1.256 against 1.291\,ms is noise.
\item \emph{Splitting kernels by match length helps.} 1.761 against
1.291\,ms, 36\% worse: thread idling is not the bottleneck.
\item \emph{The spike is a memory-geometry pathology.} Line-span 15.1 below
the 24.7 of controls; 2.1\% overlap bytes against 38.7\%; reuse 7.81 against
3.97. The cluster is instruction-dense, not memory-pathological.
\item \emph{Raising the minimum match length is a speed lever.} At high
thresholds the addressability the format exists for is lost, and the relative
spike \emph{grows}, from $2.2\times$ to $20\times$ (Figure~\ref{fig:minmatch}).
\item \emph{Our own error, corrected in the same session.} We compared a
blockwise format against whole-stream zstd. Constraints must be equalised
before measuring; Table~\ref{tab:class} is the corrected form.
\end{enumerate}

\section{Related Work}
Gompresso~\cite{gompresso} is the closest prior system and is a single
work, not two: it offers two variants (Huffman-coded and byte-coded) and two
strategies for back-reference dependencies---multi-round resolution during
decode, and dependency elimination at compression time. The latter does not
insert checkpoints; it forbids nested references that would create
intra-warp dependencies by searching only below a per-warp high-water mark.
Its reported price is up to 19\% degradation in compression ratio for that
strategy, summarised in their conclusion as no more than 10\% overall, for a
$2$--$3\times$ decode gain on a Tesla K40.

That dependency is not the one we remove. Their elimination targets the
match layer; absolute offsets and block independence already remove the
match-layer dependency in this format~\cite{aceapex1}. What
Section~\ref{sec:lastseq} removes is a \emph{parse-layer} dependency---a
four-entry distance history, absent from DEFLATE-style formats and present
in zstd. The two costs, 19\% and 0.540\%, are therefore not directly
comparable, though both are encode-time decisions paid for in ratio.

CODAG~\cite{codag} raises decode parallelism by mapping compressed chunks to
warps at runtime rather than varying compression granularity, and reports GPU
decode as compute-bound---consistent with the parse share we measure here.

Paper~4 of this series~\cite{aceapex4} identified work granularity as the
governor of decode throughput; the bus measurement in
Section~\ref{sec:floormem} arrives at the same mechanism from the traffic side.

Bounding the height of an LZ parse is studied in theory. Height-bounded
LZ~\cite{lzhb} bounds worst-case \emph{access time} through the height of
the referencing forest, giving $O(h)$ predecessor queries, and shows that
finding the smallest parsing under a height bound is NP- and APX-hard. Our
result is complementary and, on this hardware, deflationary: we provide a
practical encoder that reaches a valid parsing under the bound bit-perfectly
on a full corpus, and then measure that on a GPU the height $h$ costs almost
nothing---at most 2.8\%, and provably zero where the file's own latency
spike lives. LZ-End~\cite{kreft} addresses
extraction time from compressed text and work on balanced straight-line
programs~\cite{slp} addresses asymptotic derivation depth; both are
algorithmic statements, while ours concerns the physical realization of a
bounded-depth absolute-offset parse as a CUDA and VRAM execution layer.

\section{Limitations}
\label{sec:limits}
\begin{itemize}\itemsep2pt
\item \textbf{One GPU.} H100 80\,GB SXM. Behaviour on other generations is
untested; the 4.5--4.7\,$\mu$s wave cost is a property of the CUDA graph on
this hardware.
\item \textbf{Bench noise is 6\%} on the dense full pipe. We do not claim
effects below it. Depth is \emph{not} claimed to be irrelevant---on the
wavefront decoder it is measurable and equals $-2.6\%$.
\item \textbf{The 50\,GB seek is not bit-perfect}; the original is not on
disk. Position-invariance is measured, correctness shown on smaller corpora.
\item \textbf{Removing the distance history is not verified on GPU.} The
chain is removed and measured on CPU; the resulting decode speedup is an
\emph{estimate} and is not claimed as a result.
\item \textbf{The minimum-match probe is not bit-perfect}---tokens were
dropped without substituting literals. It illustrates a mechanism only.
\item \textbf{Wavefront numbers are match-layer only}, with entropy outside
the timer. They are never decode throughput.
\item \textbf{The class matrix covers five corpora.} The Pizza\,\&\,Chili
sets were measured on the GPU bench but are not in the ratio matrix.
\end{itemize}

\section*{Reproducibility}
Canonical corpus: chr1, UCSC hg38, md5
\texttt{9465e0f0df6e2c6eb39729c39cee5465}, 253{,}935{,}557\,B, block size
16{,}384.

Second corpus for the depth distributions: \textbf{ERR194147} (NA12878,
Illumina Platinum; ENA path \texttt{ERR194/ERR194147}), md5
\texttt{9af9ffaa\ldots}, slice of exactly
$3052 \times 16384 = 50{,}003{,}968$\,B---the first $\approx$50\,MB, not
1\,GB. A separate 50{,}000{,}000{,}000\,B slice of the same accession is the
archive of Section~\ref{sec:seek50}.

Golden corpora: enwik8, enwik9, silesia.tar, Pizza\,\&\,Chili
dna/english/proteins at 200\,MB.

Artifacts are archived at DOI \texttt{10.5281/zenodo.21874972} (release tag
\texttt{paper5-v1}, commit \texttt{22de9dc}); the concept DOI
\texttt{10.5281/zenodo.20440964} always resolves to the latest version. The
repository contains the run-parallel wavefront kernel \texttt{wf\_par.cu},
its semantics-only control \texttt{wf\_run.cu}, the encoder with the
\texttt{FORCED\_BIN}, \texttt{NO\_REP} and \texttt{MIN\_MATCH} paths used
here, and \texttt{reproduce\_paper5.sh}. MIT licensed.

\section*{Appendix: Reproducibility Contract}
Every figure in this paper carries a level rather than a caveat in prose.
\textbf{R} is reproducible: a command, an expected value and a tolerance,
with code, script and corpus fixed by accession and digest. \textbf{M} is
measured but not bit-perfect, with the reason the judge cannot be applied
stated explicitly. \textbf{E} is an estimate computed from measured
quantities, with its formula and inputs given.

\textbf{R.} Parse share 63.7--71.5\% on four corpora; depth on all three
decoders; the byte-level hash test over 15{,}499 blocks; run parallelism
$2.75$--$8.42\times$; the no-repeat cost of 0.540\%; full-pipe chr1 at
71.0\,GB/s; the class matrix of five corpora against five codecs; the
$\mathrm{ratio}(L)$ curve; the 6\% bench noise.

\textbf{M.} The 50\,GB seek---the original is not on disk, so the FNV values
have nothing to compare against. The minimum-match probe---tokens were
dropped without substituting literals.

\textbf{E.} The $2.7\times$ figure for a scan parser: parse 2.555\,ms scaled
by the 93.94\% parallelisable share gives roughly 0.15\,ms, which has not
been verified on a GPU.

\texttt{reproduce\_paper5.sh} writes one JSON record per claim---identifier,
level, expected value, tolerance, measured value, verdict, command---so that
checking the paper against a run is mechanical rather than visual. A fresh
clone of the tagged release, on an EPYC 4344P host without a GPU, reproduces
every \textbf{R} claim reachable there: \textbf{17 pass, 0 fail, 6 skipped}
(three GPU claims need a CUDA device; three are the declared \textbf{M} and
\textbf{E} entries). The recorded run ships as \texttt{results.json}.

\end{document}